# On the possibility for short time "when, where" Earthquakes prediction with the using Geomagnetic field measurements

Strachimir Chterev Mavrodiev
mavrodi@inrne.bas.bg
Institute for Nuclear Research and Nuclear Energy, Bulgarian Academy of Sciences, Sofia
August, 2003

## Abstract

This paper is an attempt for arguing the possibility for short time "when, where" and "how" Earthquakes prediction.

The local "when" Earthquake prediction is based on the connection between geomagnetic "quakes" and the next incoming minimum or maximum of tidal gravitational potential. The probability time window for the predicted earthquake is +/-1 day for the minimum and +/-2 days for the maximum. The preliminary statistic estimation on the basis of distribution of the time difference between predicted and occurred earthquake for the period 2001-2002 in Balkan, Black Sea region is given. Examples for posteriori earthquakes "when" predictions at different World regions, using archive Intermagnet local data, are given.

The possibility for local "when, where" Earthquake prediction is based on the accurate, with special space and time scales, monitoring of the electromagnetic field under, on and over Earth surface. The periodically upgraded information from Seismic hazard maps is essential.

The possibility for local "when, where and how" Earthquake prediction is based on the monitoring of geomagnetic field, electro-potential distribution in the Earth crust and atmosphere, spatial and time distribution of Earth surface radiation, gravitational anomalyties map, season and day independent temperature depth distribution, water sources parameters (debit, temperature, chemical composition, radioactivity), gas emissions, ionosphere condition parameters, Earth radiation belt parameters, Sun wind, Seismic hazard maps information for crust parameters (strain, deformation, displacement) and biological precursors.

## Introduction

The "when, where and how" earthquake prediction is not solved problem [1-9].

The more precision space and time set for Earth's crust condition parameters and the including in the monitoring the electromagnetic fields measurements under, on and over Earth surface, the temperature distribution and other possible precursors can be useful for research of the "When, where and how" earthquake's prediction [10-22]. The progress in this direction partly is summarized in [9-22]. The approach is based on the understanding that earthquake processes are part of very complex and without adequate physical model of Earth existence and the gravitational and electromagnetic interactions, which en sure the stability of Sun system and its Planets for long time. The earthquake part of the model can be recovered in the infinity way "theory- experiment-theory" using nonlinear inverse problem methods for looking the correlations between fields in dynamically space and time scales.

The achievements of tidal potential modeling of Earth surface with included ocean and atmosphere tidal influences is essential part of the system.

The today almost real time technologies GIS for archiving, visualization, analysis and interpretation of the data and non- linear inverse problem methods for building theoretical models for the parameter behaviors, correlations and dynamics have to be used.

Multiple component correlation analysis and the nonlinear inverse problem digital and analytical methods in the frame work of fluids dynamics and Maxwell theory have to be crucial for the building step by step the adequate physical model.

The role of geomagnetic variations as precursor can be explained by the hypothesis that in the time of earthquakes preparing, with grows of strain, deformation or displacement in the Earth depth in some interval of density changing, arises the chemical phase shift which leads to an electrical charge shift. The arrived new system of alternating electrical currents with periods in minute scales is the origin of magnetic field, which contribution to the geomagnetic field is geomagnetic "quake". The K-index can not indicate the local geomagnetic variation, because of the hour scale calculation. Nevertheless, the K- index behavior in the near space has to be analyzed. The second moment of magnetic field, calculated for minute time scale, can indicate a bigger, than usual, variations for seconds and minutes scales. If the field vector components are measuring many times per second, one can calculate the frequency dependence of full geomagnetic intensity and to analyze the frequency spectrum of geomagnetic quake. If the variations are bigger for some time, than usual, one can say that we have the geomagnetic quake - the earthquake precursor. The probability time window for the predicted earthquake (event, events) is approximately +/-1 day for the minimum and +/-2 days for the maximum of Earth tidal potential behavior.

The future epicenter coordinates can be estimated on the basis of geomagnetic vector measuring at last in three points and with the inverse problem methods, applied for estimation the coordinate of the volume, where the phase shift arrived and its time window.

In the case of incoming big earthquake (Magnitude > 5 - 6) the changes of vertical electropotential distribution, the Earth temperature, the infrared Earth radiation, the behavior of water sources, its chemistry and radioactivity, the atmosphere conditions (earthquakes clouds, etc.), the charge density of Earth radiation belt have to be dramatically changed in and over epicenter area.

The achievements of tidal potential modeling of Earth surface with included ocean and atmosphere tidal influences, many component correlation analysis and nonlinear inverse problem methods in fluids dynamics and electrodynamics are crucial for every step of building the mathematical and physical models.

The today real time communication and geographical information system for archiving, visualization, analysis, and interpretation of data and free Internet publication is essential part of the system, which will permit the including in the science group new scientists with different interests.

In Part 1 are given the 2002 statistics estimations for the reliability of the time window earthquake prediction on the basis of geomagnetic field measurements and Earth tidal behavior [25] for Balkan, Black Sea region [17].

In Part 2 the posteriori analysis of the approach is applied for Alaska region 2002 Mag 8.2 earthquake and Hokkaido, 2003.

In Part 3 the list of monitoring parameters, which can be useful, the theoretical apparatus for analysis and technology for data acquisition are described shortly

## Part 1.The geomagnetic field quake likes a time window earthquake's precursor for Balkan, Black Sea region

It is useful to stress that the author's interests to the earthquake's prediction problem arrived as a result of complex research of the Black Sea ecosystem some 15 years ago [23]. In the time of gathering the historical data for the ecosystem arrived the facts for the Crime earthquake in 1928, which should be interpreted as electromagnetic and earthquake correlations. The Russian scientist academician Popov proposed such hypnotizes in the early 30-th of 20-th century- private communication.

Toward Intermagnet requirements for measuring the geomagnetic field (see Figure 1) on Earth surface [30], the accuracy is ±10 nT for 95% of reported data and ±5 nT for definitive data, with one sample per 5 seconds, in the case of Vector magnetometer ($\mathbf{F}$(XYZ) or $\mathbf{F}$(HDZ)) and 1 nT, with 3 samples per second, for Scalar Magnetometer ($\mathbf{F}$).

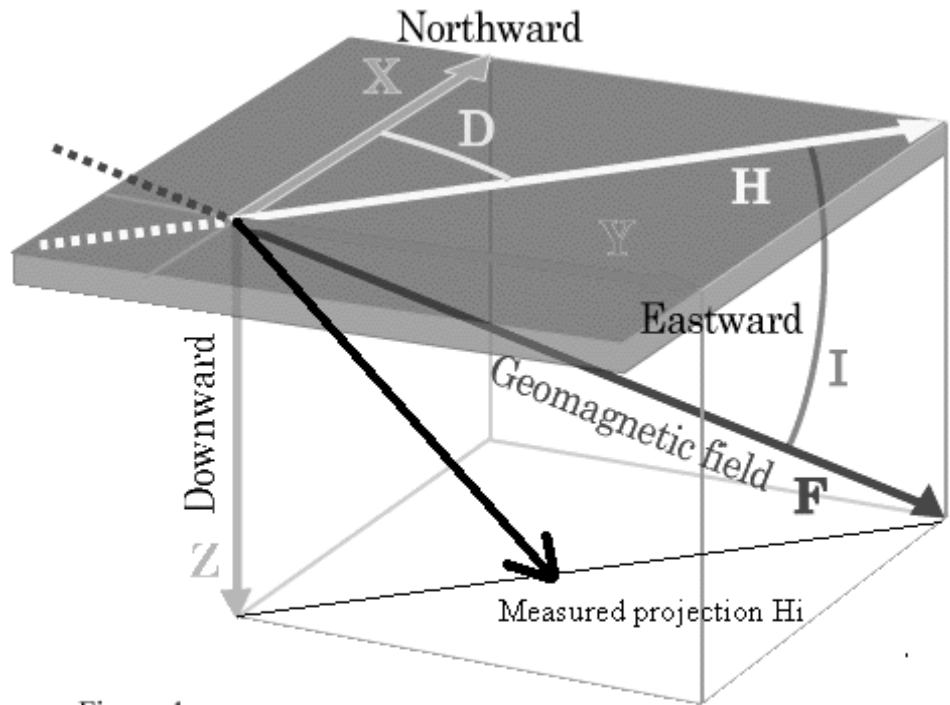

Figure 1
The components of Earth Geomagnetic field
Source: http://swdcwww.kugi.kyoto-u.ac.jp/element/eleexp.html

The geomagnetic vector projection H is measured with accuracy less or equal to1 nT (know-how of JINR, Dubna, Boris Vasiliev) with 2.4 samples per second. Because of technical reason the sensor was oriented under the Horizon in such manner that the measured value of $H_i$ is around 20000 nT. See Figures 1 and 2.

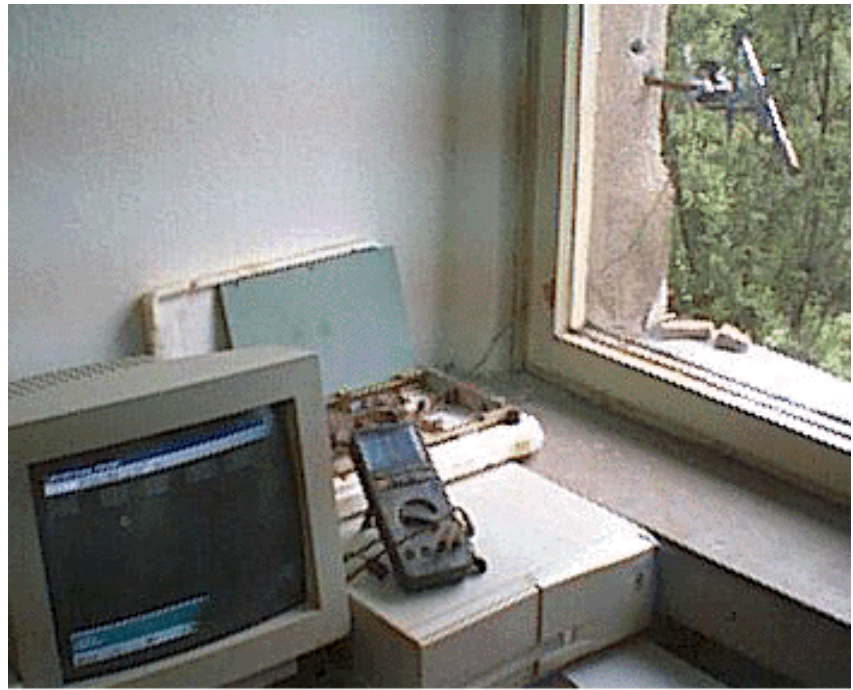

Figure 2 The Magnetometer with Sensor and PC

The minute averaged value $H_m$ and its error $\Delta \bar{H}_m$ are

$$\overline{H}_m = \sum_{i=1}^{N_m} \frac{H_i}{N_m} \text{ and } \Delta\overline{H}_m = \sum_{i=1}^{N_m} \frac{\Delta H_i}{N_m},$$

where $H_i$, $\Delta H_i$ are the measured Nm=144 times per values of the field and their experimental error. The standard deviation $s_{H_m}$ and its error $\Delta s_{H_m}$ for every minute are

$$s_{H_m} = \frac{\sqrt{\sum_{i=1}^{N_m}(H_i - \overline{H_m})^2}}{N_m} \text{ and } \Delta s_{H_m} = \frac{\sqrt{\sum_{i=1}^{N_m}(\Delta H_i - \overline{\Delta H_m})^2}}{N_m}$$

.

After some time (from 1999 to 2001) of looking for correlations between the behavior of the geomagnetic field, Earth tidal gravitational potential and the occurred earthquakes one turn out that the daily averaged value of $s_{H_m}$ and $s_{\Delta H_m}$, which we denote by Sig (ΔSig), is playing the role of earthquake precursor.

The Figure 3 illustrates the behavior of geomagnetic field component $H_m$ and its variation $s_{H_m}$ for a day without signal for near future "big and near enough" earthquake in the region.

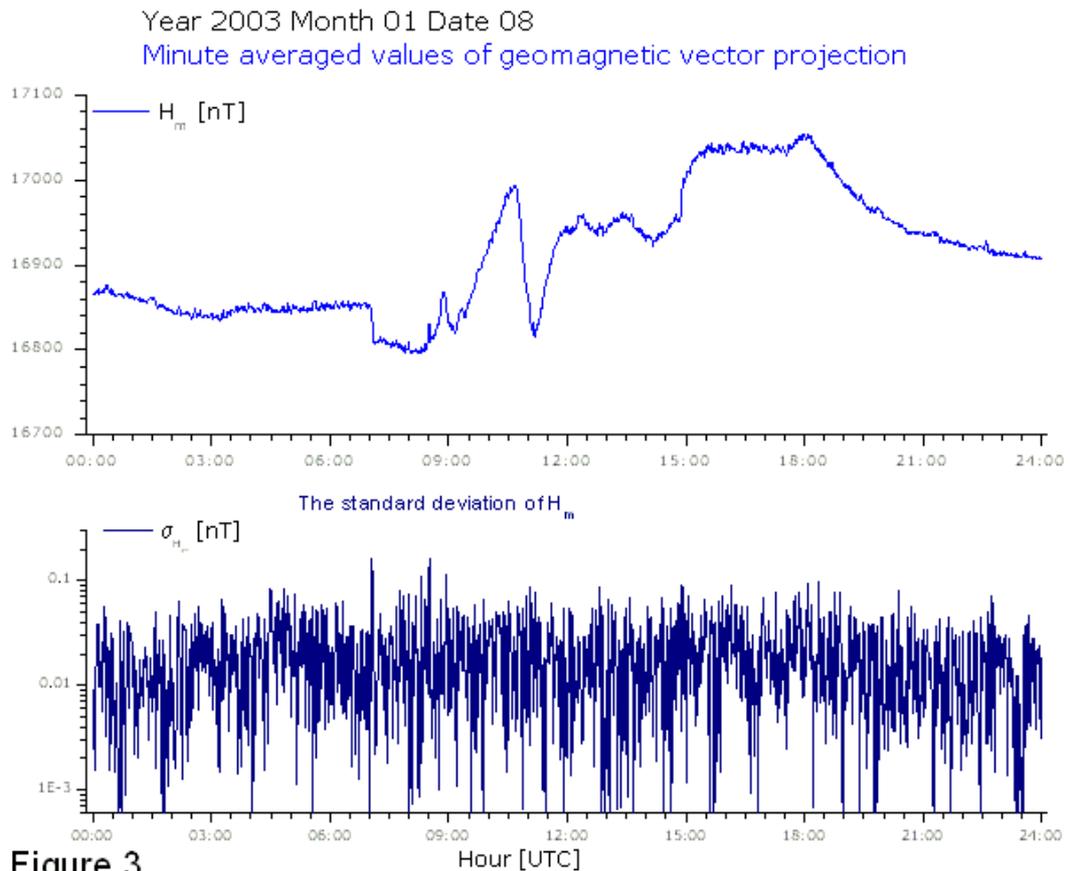

Figure 3
The behavior of geomagnetic field for a normal day

The Figures 4, 5 illustrate the behavior of geomagnetic field and its variation for a day with a signal for near future earthquake. One has to be sure there are not a cosmos or Sun wind reasons for the geomagnetic quake. So, for example, see the sites [27-29].

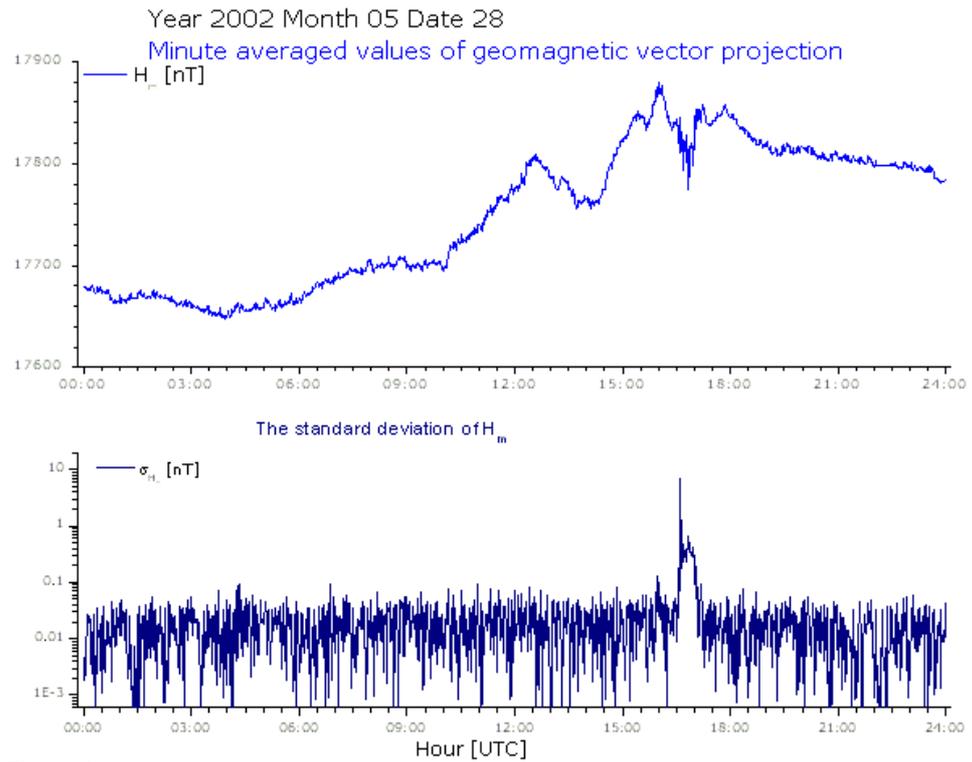

Figure 4
The behavior of geomagnetic field for a day with a signal for a near future earthquake

For the example in Figure 4 the predicted time was 3 +/- 1 June, 2002 and was confirmed with earthquake, occurred at 03/06/2002 02:04, Lat41.95N, Lon23.10E, Dep8, Mag2.6, Ml, 50 km from Sofia, $S_{ChtM}$= 598 [Mag/r$^2$].

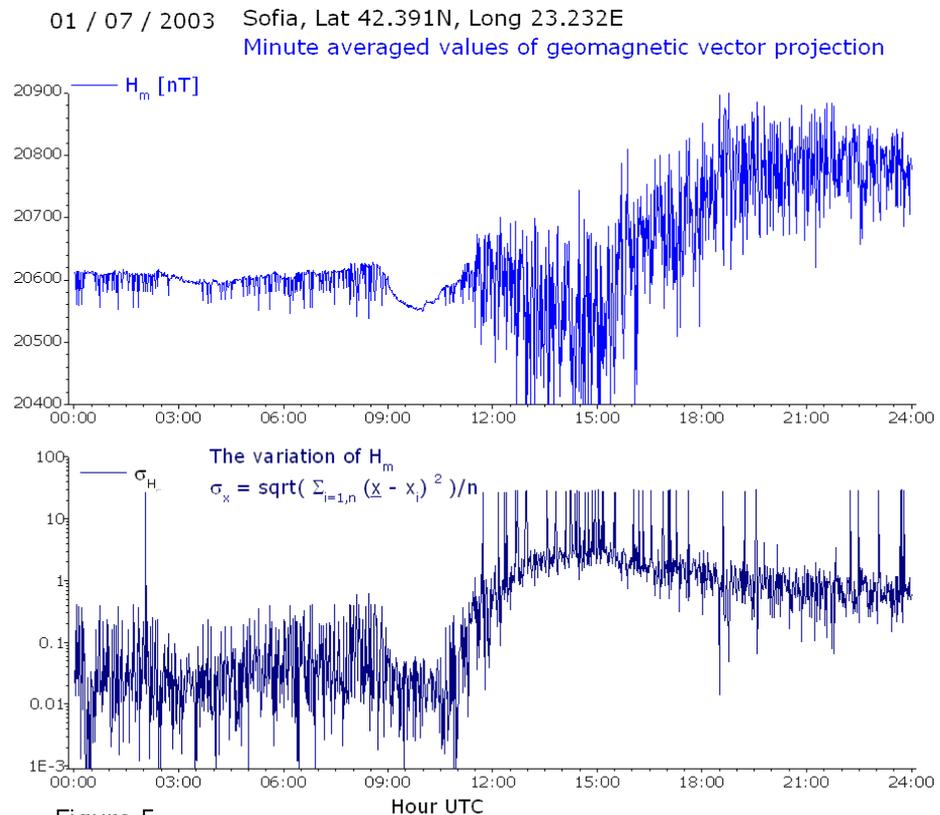

Figure 5
A day with geomagnetic quake, which is eq's precursor

The preliminary Fourier analysis of $H_m$ data give that the bigger geomagnetic variations are cased from the arriving for hours time period a new frequencies with periods from 10-th of seconds till 10 minutes and with very specific amplitude behavior. Such new specter arriving for hour time is invisible for minute samples measuring.

The probability time window of the incoming event or events is defined by the next date of the Earth tidal potential extremum with tolerance +/-1 day in the case of minimum and +/-2 days in the case of maximum. See Figure 6.

The uncertainty problem of distinguishing the predicted event (or group of events – aftershocks) from the events which occurred in the region in the predicted time window is solved with new earthquake function $S_{ChtM}$:

$$S_{ChtM} = \frac{2Magnitude}{(R_{eq} + Dis\tan ce)^2}, R_{eq} = 0.040 + \frac{Dept}{Magnitude} \text{ [Thousand km]}$$

The sense of function $S_{ChtM}$ is a density distribution of earthquake's Magnitude. In the point of measurement $S_{ChtM}$ is logarithmically proportional to the energy influence of the earthquakes. It is important to stress out that the first consideration of the Magnitude and distance dependences was obtained on the basis of nonlinear inverse problem methods. Obviously, the nearer and biggest earthquake (relatively biggest value of $S_{ChtM}$) will bear more electropotential variations, which will generate more power geomagnetic quake. At this stage of the researching the measure of daily geomagnetic state can serve the averaged for 24 hours (1440 minutes) values of standard deviations:

$$(1) \qquad Sig = \sum_{i=1}^{1440} \frac{\boldsymbol{s}_{H_m}}{1440}, \ \Delta Sig = \sum_{i=1}^{1440} \frac{\Delta \boldsymbol{s}_{H_m}}{1440}$$

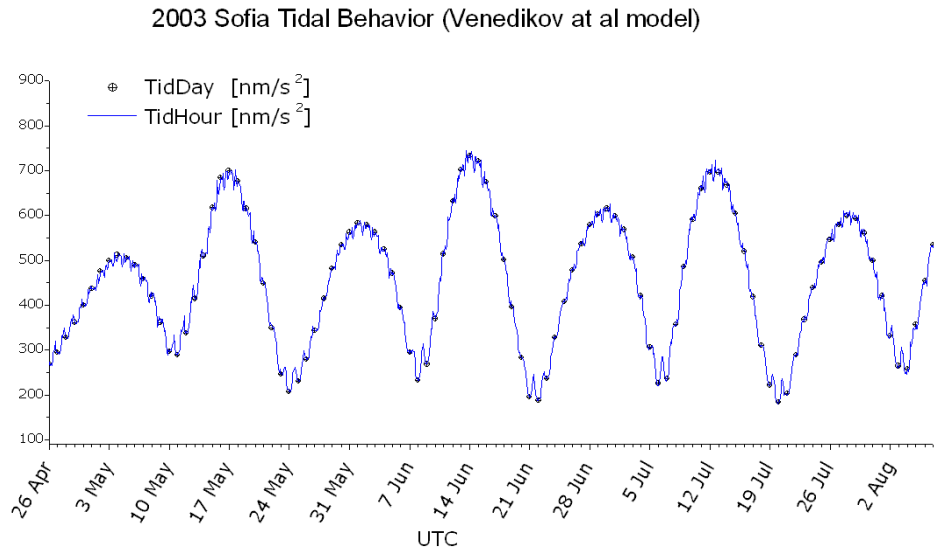

Figure 6
The Tidal Gravitational Potencial behavior for June, 2003
Vinedikov at al model[ ]

**The simple and usually working criteria for evidence of geomagnetic quake is when Sig increase for two consecutive days i, i+1 and the difference between the values of Sig is bigger than the mean arithmetic sum of their errors DSig:**

$$(2) \qquad Sig_{i+1} - Sig_i \ > \ (DSig_i + DSig_{i+1})/2 \ , \ Sig_{i+1} > Sig_i$$

If criteria (2) are fulfilled and there are not a cosmos and Sun generated variations of geomagnetic field, one can say that the geomagnetic quake had a place. Such quake is unique precursor for incoming earthquake and in the next minimum or maximum of the local Tidal gravitational potential somewhere in the region this predicted earthquake will occur.

For some of the cases, the criteria (3) have to be calculated for hour or ten's of minute's time scale.

The signal for earthquakes with different epicenters we have in the case when the specific behavior of field and its standard deviation occur more than one time at different hours of the day.

The analysis of the precursor function Sig on the basis of special kind 5 points digital derivatives can serve in the future for creating the algorithm for automated alert system.

Obviously, the more detailed time window can be achieved with analyze of daily variations of tidal potential, calculated every hour.

For the concrete example in Figure 5 the predicted events had occur with parameters:

| DDMMYYhhmm | Lat | Lon | Dep | Mag | Dist | SChtM |
|---|---|---|---|---|---|---|
| 03/07/2003 15:49 | 42.96 | 25.29 | 10 | 2.6 | 1.10 | 219 |
| 03/07/2003 20:51 | 41.96 | 23.28 | 12 | 2.9 | 0.48 | 684 |
| 05/07/2003 21:58 | 40.35 | 26.14 | 2 | 4.0 | 2.97 | 70 |
| 06/07/2003 19:10 | 40.46 | 26.01 | 10 | 5.7 | 3.16 | 89 |
| 06/07/2003 19:39 | 40.62 | 25.25 | 2 | 4.2 | 2.35 | 111 |
| 06/07/2003 20:02 | 40.72 | 25.98 | 20 | 3.2 | 2.54 | 71 |
| 06/07/2003 20:10 | 40.46 | 26.08 | 10 | 5.0 | 3.20 | 76 |
| 06/07/2003 20:48 | 40.28 | 26.08 | 2 | 4.1 | 3.01 | 70 |
| 06/07/2003 21:58 | 40.38 | 26.10 | 10 | 3.9 | 2.92 | 70 |
| 06/07/2003 22:05 | 40.34 | 26.00 | 2 | 3.9 | 2.92 | 71 |
| 06/07/2003 22:42 | 40.95 | 26.00 | 10 | 4.6 | 2.80 | 89 |
| 07/07/2003 00:24 | 40.25 | 25.98 | 2 | 3.7 | 3.00 | 64 |
| 07/07/2003 00:48 | 40.44 | 25.87 | 18 | 3.4 | 2.77 | 66 |
| 07/07/2003 07:15 | 41.67 | 24.88 | 10 | 3.1 | 1.21 | 230 |
| 07/07/2003 16:17 | 40.37 | 25.91 | 10 | 3.3 | 2.85 | 61 |
| 08/07/2003 02:48 | 41.82 | 22.93 | 12 | 2.8 | 0.65 | 465 |
| 08/07/2003 12:00 | 42.84 | 23.32 | 10 | 2.5 | 0.50 | 565. |

At this stage of research all earthquakes have the same $S_{ChtM}$ for different definitions of the Magnitude. After developing on the basis of inverse nonlinear problem the mathematical models of empirical and theoretical dependences between incoming earthquake processes, magnetic quake and parameters of earthquake we will arrive to a set of $S_{ChtM}$ functions in correspondence with the different definition of Magnitude. The volumes, its depth, chemical and geological structures of the region have to be included in the dependences.

It is interesting to stress that in the case of big earthquake with Mag > 6, in more than 60 % of the cases, after the earthquake there are very little variations, compared with usual daily behavior, of the magnetic field with different time duration: from 10 minute to some hours. Those influences do not depend on the distances, but one can see the differences, which depend on the zone of earthquake: convergence or divergence one. See for example next Figure 7.

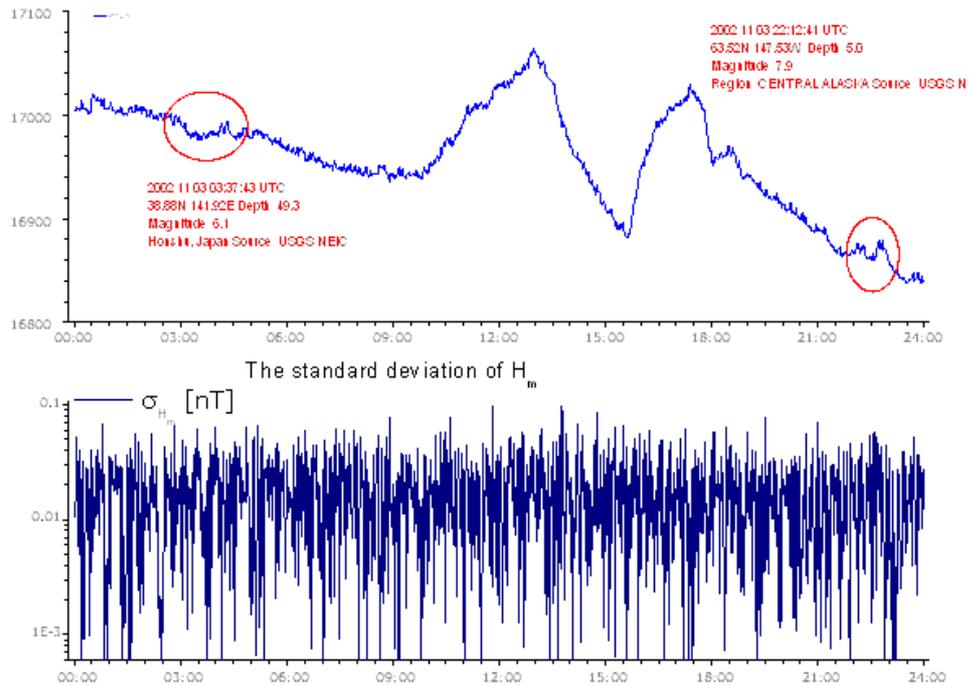

Figure 7
A registration of big world earthquake, Mag>6

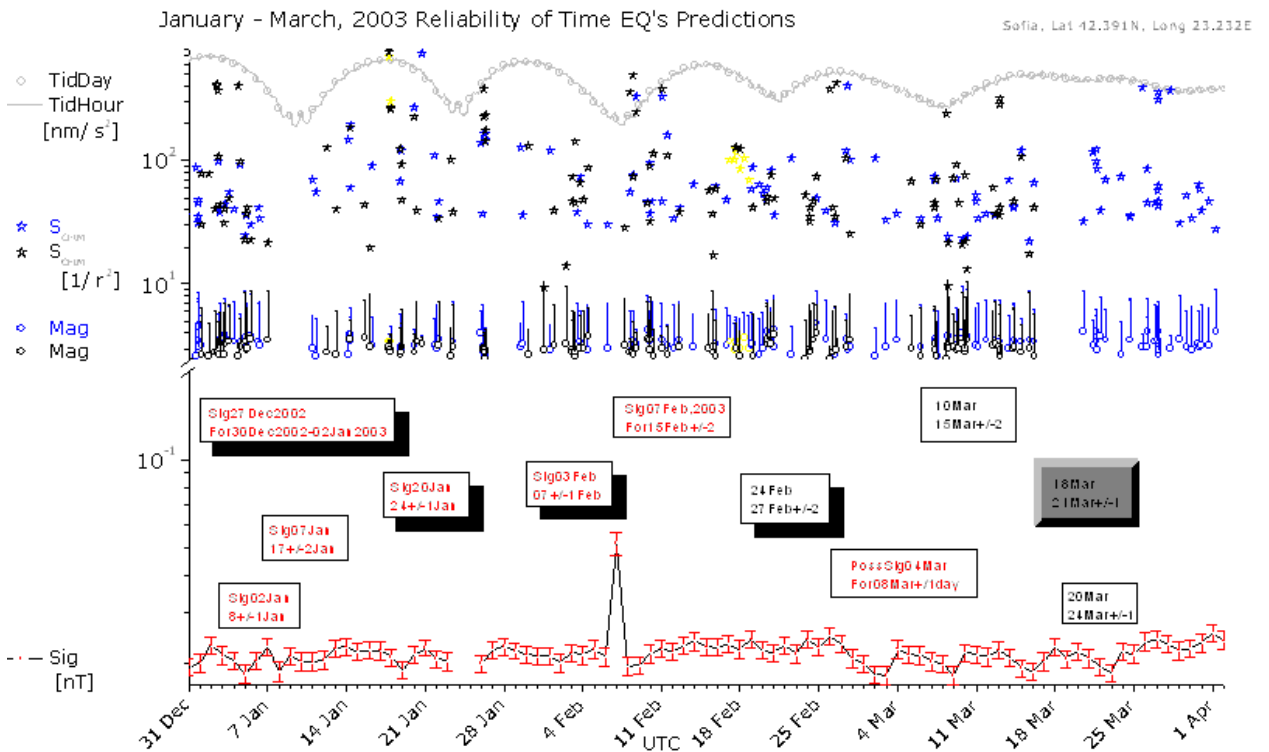

$S_{ChtM} = 2Mag/(R_{eq} + Distance)^2$, $R_{eq} = 0.040 + Dept/Mag$ [thousand km]

$Sig = \sum_{m=1,1440} \sigma_H /1440$, $\sigma_H = sqrt( \sum_{i=1,144} (H_i - \underline{H}_m)^2)/144$, $\underline{H}_m = \sum_{i=1,144} H_i /144$,

where $H_i$ is the geomagnetic vector projection, measured with 2.4 samples per second.

The vertical error of Mag is distance [Hundred km].The precursor is Sig irregularity.The predicted time is defined from the next Tidal potentiaal(Venedikov et al model) minimum (+/-1day) or maximum (+/-2). Data from http://wwwneic.cr.usgs.gov/neis/bulletin/, GPhI,BAS,Sofia

Figure 8

In Figures 8 and 9 at the same graphic the Tidal potential, $S_{ChtM}$, Magnitude, Distance from Sofia, Sig are represented. In the text boxes are the data of the geomagnetic quake (date of the precursory signal) and the time window for predicted event (events).

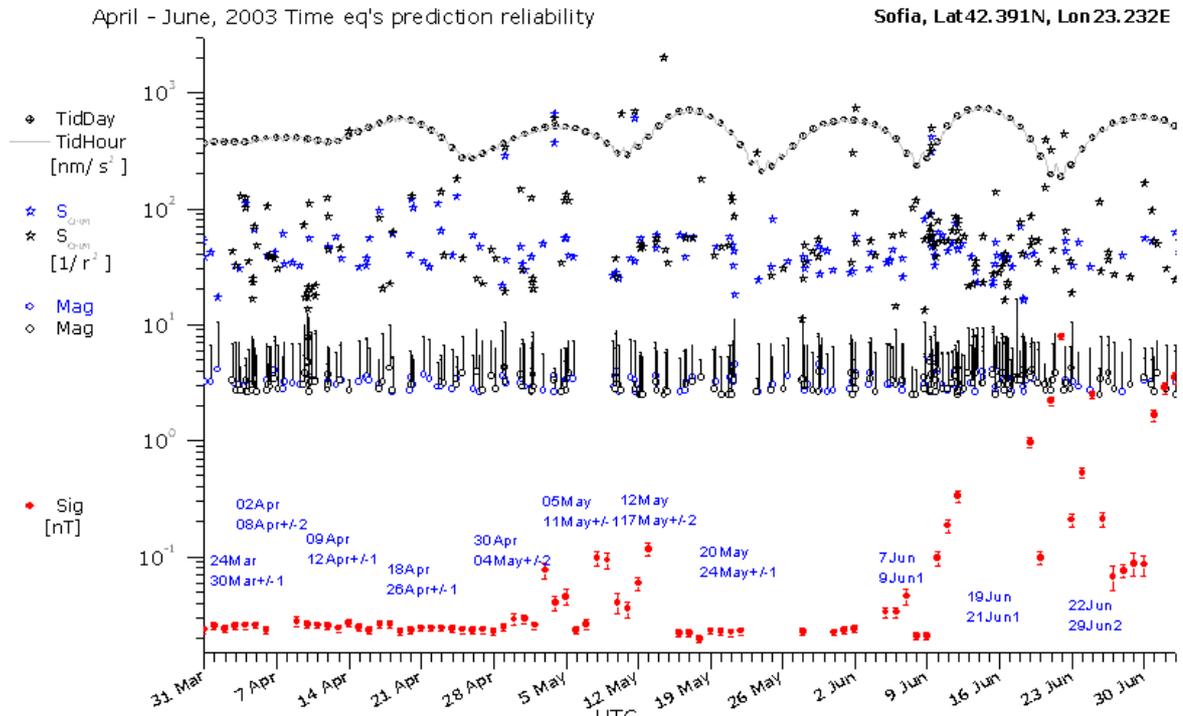

$$S_{ChtM}=2Mag/(R_{eq}+Distance)^2, \quad R_{eq}=0.040+Dept/Mag \text{ [thousand km]}$$
$$Sig = \Sigma_{m=1,1440} \, \sigma_H /1440, \quad \sigma_H =sqrt( \, \Sigma_{i=1,144} \, (H_i - \underline{H}_m)^2)/144, \quad \underline{H}_m = \Sigma_{i=1,144} H_i /144,$$

where $H_i$ is the geomagnetic vector projection, measured with 2.4 samples per second. The vertical error of Mag is distance [Hundred km].The precursor is Sig irregularity.The predicted time is defined from the next Tidal potentiaal(Venedikov et al model) minimum (+/-1day) or maximum (+/-2). Data from http://wwwneic.cr.usgs.gov/neis/bulletin/, GPhI,BAS,Sofia

Figure 9

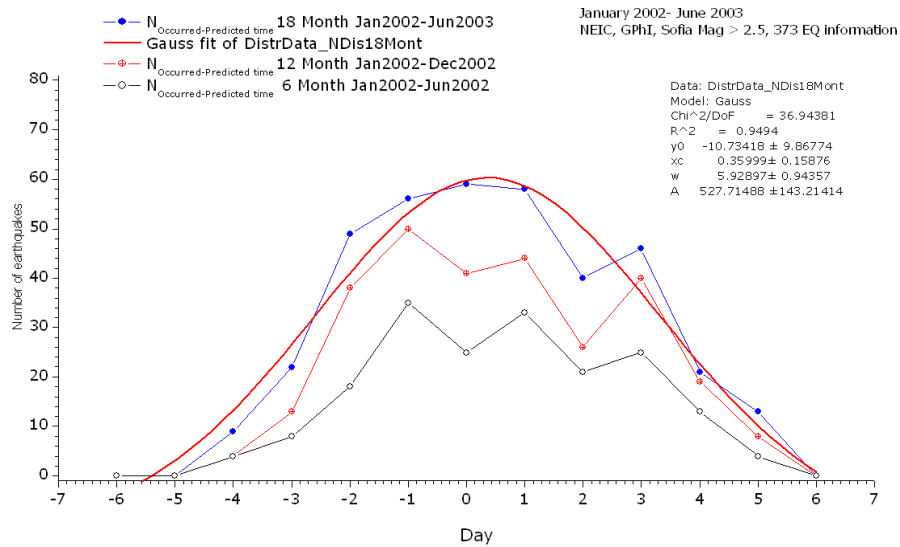

Figure 10
The distribution of difference occured - predicted earthquakes time for different time periods

In Figure 10 the distributions of the difference between times of predicted and occurred events, calculated at 6, 12 and 18 month period, starting from January, 2002 are presented. The growth of the distribution without widening and its approximation the Gauss distribution in time is argument that the correlation geomagnetic signal- Tidal potential extremum and occurred earthquake has a physical causality origin. The number of earthquakes, analyzed in Figure 10 is greater than the predictions of the events. The obvious reason is that in the case of some earthquakes with greater Magnitude there were aftershocks

On the next Figure 11 the reader can see that at distance 200 km the geomagnetic quake can stay as an earthquake precursor for earthquakes with Magnitude > 2.5.  Obviously, in greater distances the predicted events will have greater Magnitude.  The today estimation is that big earthquakes (Magnitude >5) could be predicted till distances 500- 600 km.

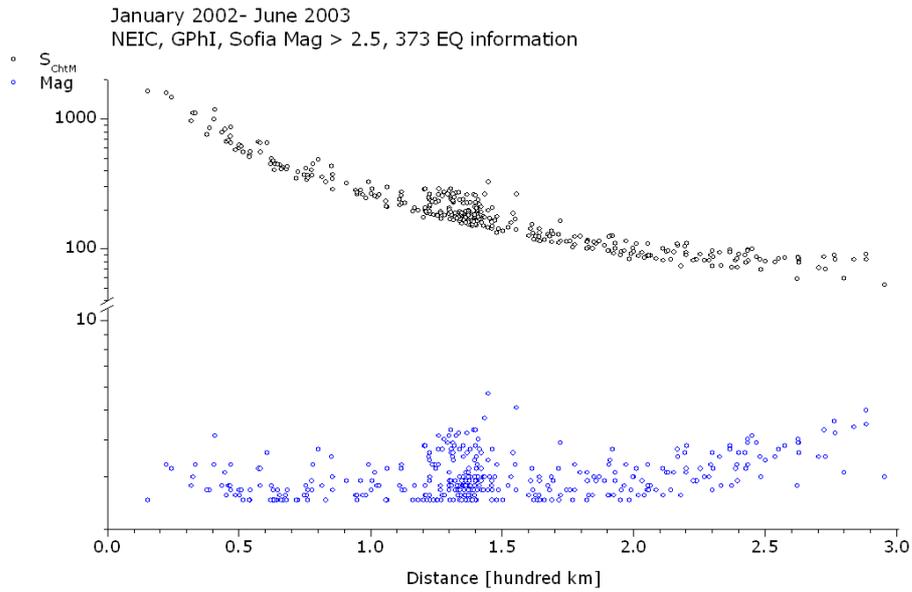

Figure 11
The $S_{ChM}$ dependence on distance

At the next Figure 12, for completeness, is represented the Magnitude distribution of predicted events.

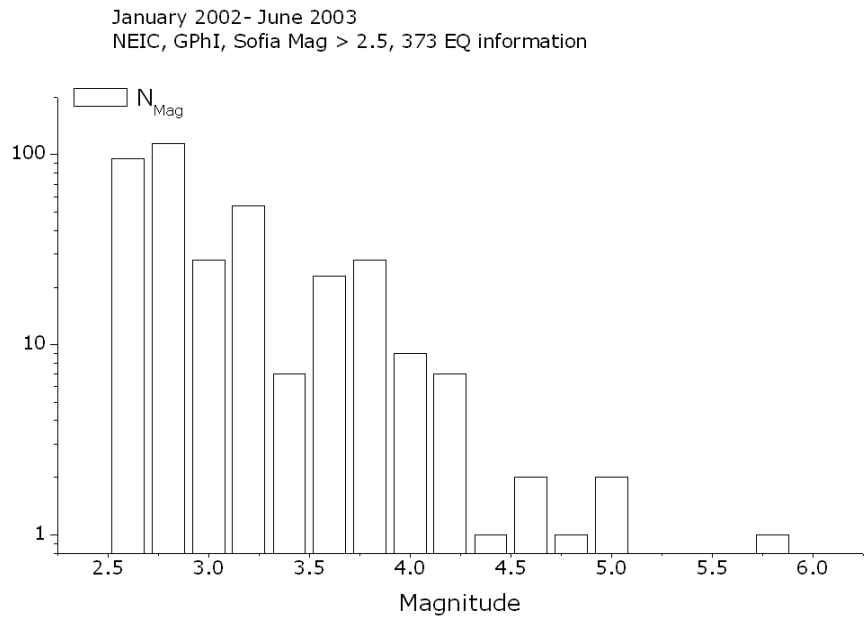

Figure 12
The EQ's Magnitude distribution

The independent control (In the framework of Strasbourg recommendations about earthquake prediction of the European Union for ethical and public security reasons) of the time window earthquake prediction reliability was organized in the framework of the Bulgarian Academy of Sciences, its Geophysical Institute and a set of colleagues, which are interested in the research topic [16- 17], starting from January 2002.

Summarizing, one can tell that for the analyzed period have occurred 69 events. For 59 of them the time window was successfully predicted. For 4 of occurred events the precursor quake was not established by the everyday analysis and 6 events were not predicted because of hardware problem.

The above results for successful and reliability time window earthquake prediction can be considerate as a fist step for solution of "when, where and how" earthquake prediction at level "when". Because of absence the adequate physical model of the correlation and theoretical estimation of the measured experimental values the possibility that the earthquake with magnitude greater than 2.5 are occurring very scare in the region to fulfill the established correlation is not closed.

The first prove, that in the framework of such complex approach, the "when, where and how" earthquake prediction problem can be solve will be the "when, where" prediction on the basis of at least 3 point electromagnetic real time monitoring. If the statistic estimation will be successful for a long time period and the established correlations will be confirmed from the adequate physical model solutions, one could say the earthquake prediction problem is under solving.

### Part 2. The posteriori analysis
### Alaska, 2002 Magnitude 8.2 earthquake, Second CMO geomagnetic data

In the case of geomagnetic vector measurements the precursor signal $Sig_F$ is defined as daily averaged sum of normalized standard deviations $\dfrac{S_{H_m}}{H_m}, \dfrac{S_{D_m}}{D_m}, \dfrac{S_{Z_m}}{Z_m}$ - see Figure 13.

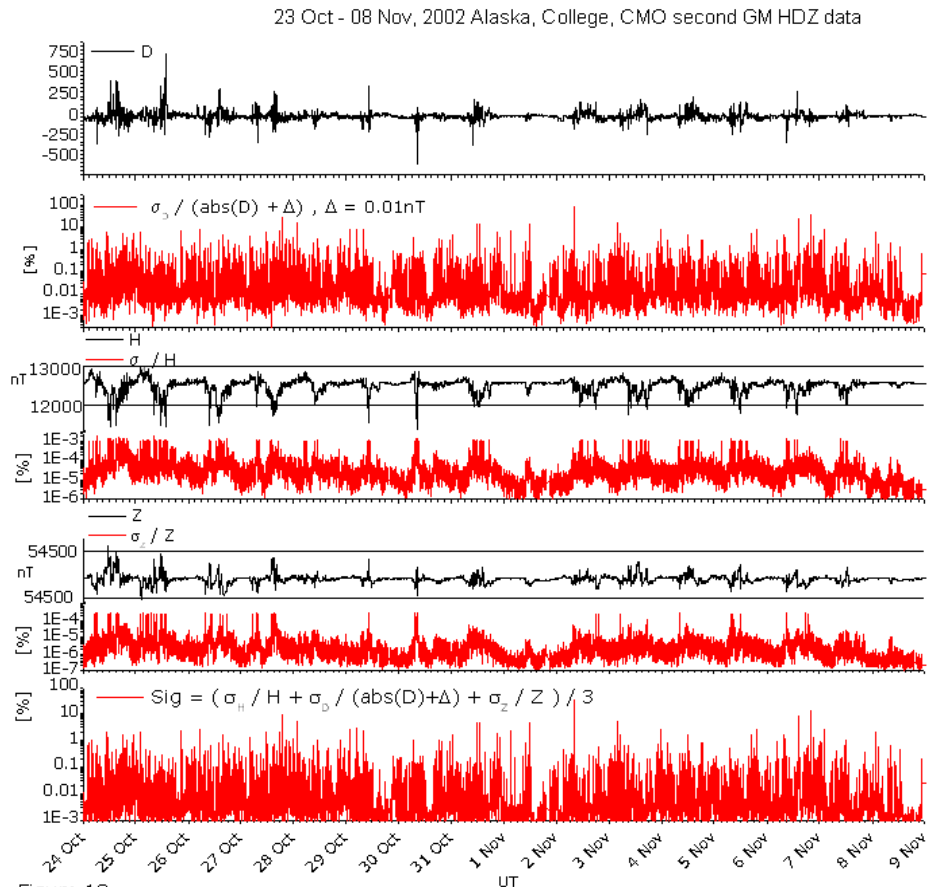

Figure 13
The minute avereged second HDZ and its normed standard deviation

In the next Figure 14 the data for Tidal daily behavior, function $S_{ChtM}$, the Magnitude of occurred earthquakes and precursor signal $Sig_F$ are presented for the period from 24 October to 9 November, 2002. The precursor $Sig_F$ is calculated using the second HDZ College Geomagnetic observatory data (CMO Intermagnet geomagnetic observatory with coordinates 64.84N, 148.86W.

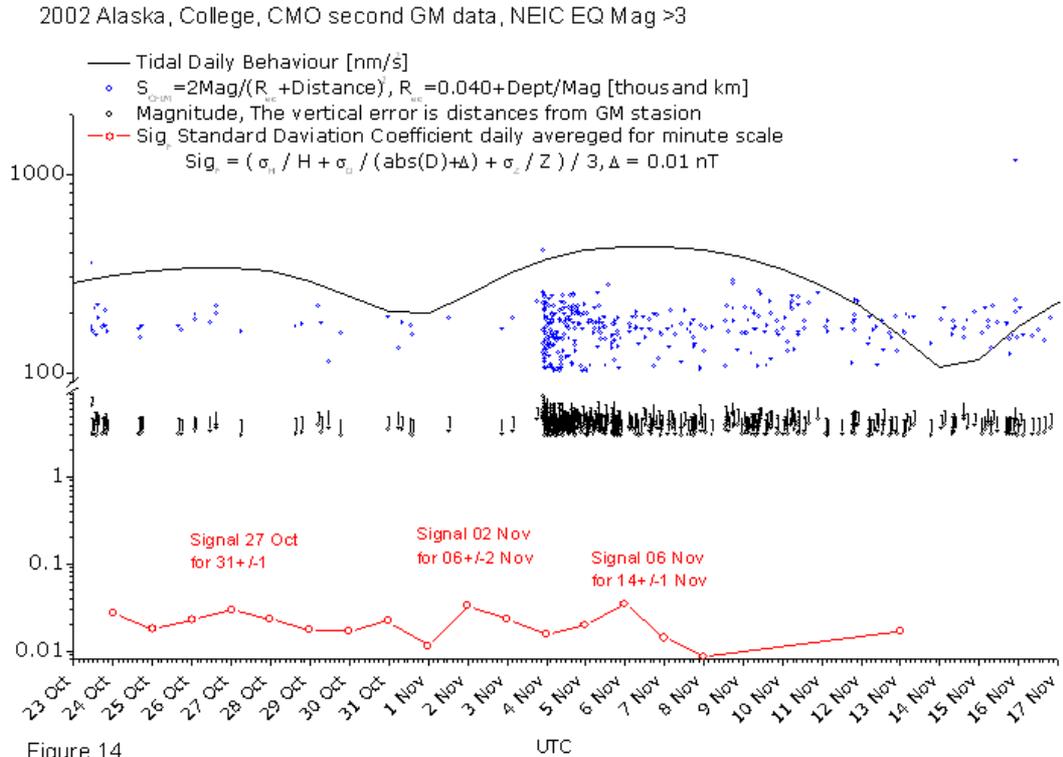

Figure 14

The dependences between Geomagnetic variations ($Sig_F$), SChtM eq parameter and Tidal potential

Although the geomagnetic data are only seconds, the correlation between geomagnetic precursor function and incoming earthquake is seen. The second CMO data was kindly given by USA Intermagnet group.

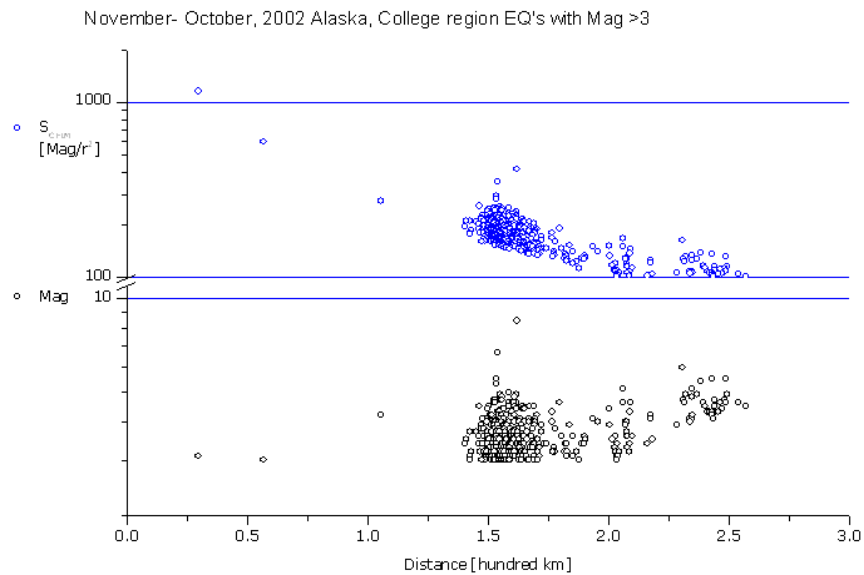

Figure 15

The distance dependences of function $S_{ChtM}=2Mag/(R_{eq}+Distance)^2$, $R_{eq}=0.040+Dept/Mag$ [thousand km]

### Hokkaido 2003, MMB minute geomagnetic data

In the next Figure 16 the data for Tidal daily behavior, function $S_{ChtM}$, the Magnitude of occurred earthquakes and precursor signal $Sig_F$ are presented for the January - June, 2003 for Hokkaido, Japan. The geomagnetic data are minute one from Intermagnet MMB observatory near Mamambetsu, 43,91N, 155.19E. The minute averaging is changing with hour one.

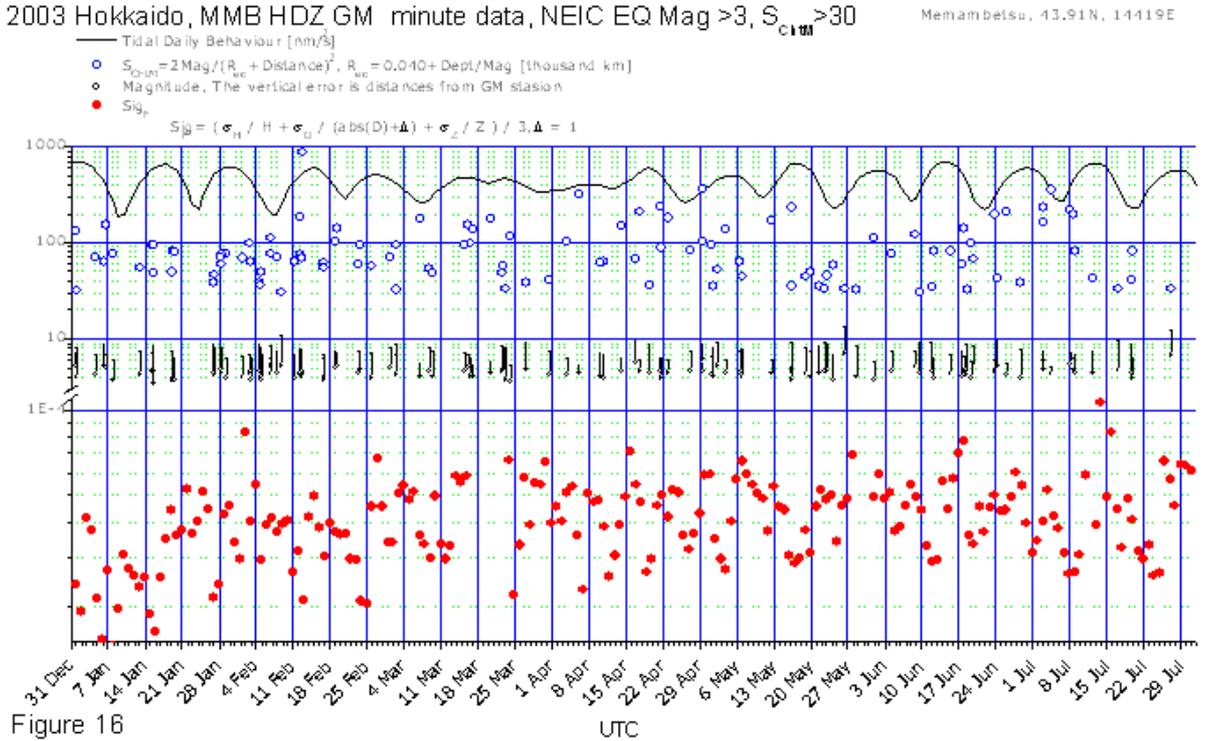

Figure 16
The Geomagnetic variations $Sig_F$ (minute data),
$S_{ChtM}$ eq parameter and Tidal potential

It is seen that there are more signals $Sig_F$ for geomagnetic quake than a strong earthquakes with big values of $S_{ChtM}$. The reason is that the standard deviation was calculated for 1 hour. Ergo, the daily variations, which are not consequences of the local geomagnetic quake, can give the analogous effect.

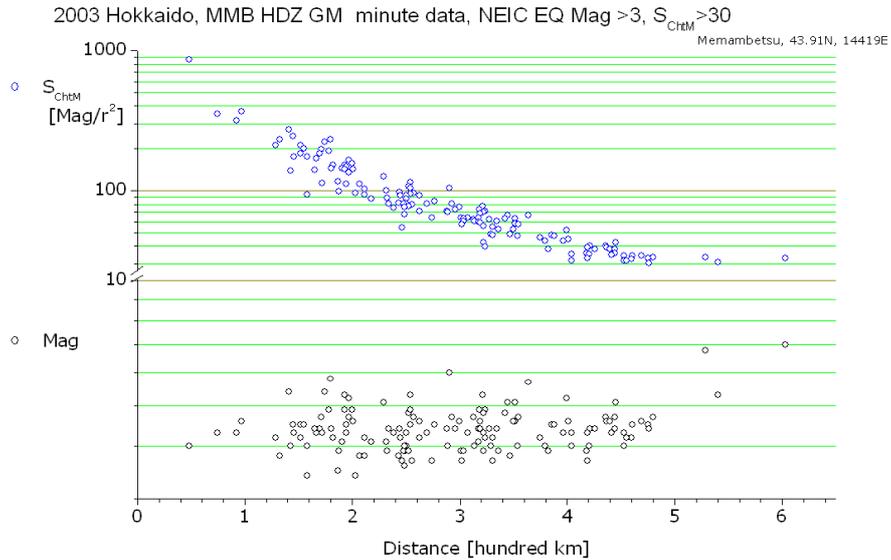

Figure 17
The distance $S_{ChtM}$ dependence

The preliminary analysis of the second MMB Intermagnet data shows that the Hokkaido could be a good polygon for testing the possibility for "when, where" eq's prediction if there is another two Intermagnet geomagnetic observatory **F**(HDZ) with at least 10 samples per second.

Analogous analysis for the regions, where are the Intermagnet geomagnetic stations, for establish the correlation between local geomagnetic quake and incoming in time window, defined from the time of next minimum or maximum of the Tidal behavior, earthquakes was performed for England, India, Turkey [21].

**Part 3. Proposal for creating of Short Time Earthquake prediction local NETWORK**

We will not discuss the long time prognostic system for estimation of earthquake risk. They are well known [32, 32].

The aim of this Proposal is to create a system for research the reliability of the local forecast system for earthquakes in the interval Mag > 2.6- 3 and radius till 600 km. The system is complex and the attended practical result will be a short time "when, where" eq's prediction. The "when, where and how" problem will be solved step by step in the building an adequate physical theoretical model for the Earth magnetism. The new type of science unification has to be realized for such complex research. The system includes experimental, theoretical and technological parts:

**Experimental data for:**
Geomagnetic field, Electro-potential distribution in the Earth crust and atmosphere, Temperature Earth crust distribution, Crust parameters (strain, deformation, displacement), Gravitational anomaly map, Season and day independent depth temperature distribution, Water sources parameters (debit, temperature, chemical composition, radioactivity), Gas emissions, Ionosphere condition parameters, Infrared radiation of Earth surface, earthquake clouds, Earth radiation belt, Sun wind, Biological precursors.

**Theory**: The achievements of tidal potential modeling of Earth surface with included ocean and atmosphere tidal influences, many component correlation analyses, Nonlinear inverse problem methods in fluids dynamics and Maxwell equations are crucial.

**Technologies: GIS** for archiving, analysis, visualization and interpretation of the data and non-linear inverse problem methods for building theoretical models for the parameter behaviors, correlations and dynamics.

The set of the devices has to be in correspondence with known data for earthquakes risk zone (gravitational anomalyties and Crust parameters monitoring (strain, deformation, displacement). The geomagnetic device set distance has to be in order of 150 – 200 km, the electro-potential 100- 200 km in dependence of geological today situation and its history. The set for monitoring of the daily and season crust temperatures has to be in order 300 km. The correlations with Sun wind influence have to be in real time.

The system has to be created step by step. The condition for next step pass has to be the building of physically clear new theoretical correlations or dynamical models and, of course, the successful: "when", "when, where" or "when, where and how" earthquake prediction.

In conclusion in Figure 18 is represented the number of world earthquake's with Magnitude greater then 4. The correlation between the eq's number, global warming and consequent increasing the Sea level and the amplitude of Ocean tides is going to be obvious.

**Lyric digression**: The total slime of Burgas bay bottom at 1996 accidentally coincide with the jump of eq's number.

**Conclusion**

The correlations between local geomagnetic quake and incoming earthquakes, which occur in the time window defined from the next minimum (+/- 1 day) or maximum (+/-2 days) of the Earth Tidal gravitational potential is tested statistically. The distribution of the time difference between predicted and occurred events is a Gauss one and is increase in the time.

The hypothesis for possible physical model is proposed.

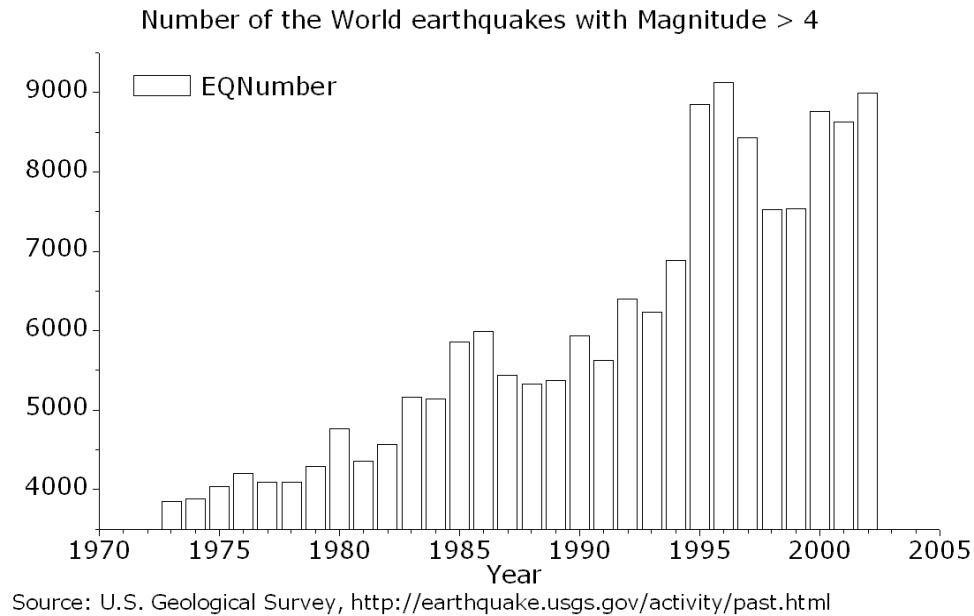

Number of the World earthquakes with Magnitude > 4

Source: U.S. Geological Survey, http://earthquake.usgs.gov/activity/past.html

Figure 18
The number of earthquake's with Magnitude > 4

This result can be interpreted like a possible first reliable approach for solving the "when" earthquakes prediction problem.

On the basis of electromagnetic monitoring under, on and over Earth surface is proposed research for solution of "when, where" earthquake prediction problem. Under the hypothesis the current has a big vertical component the data of two geomagnetic vector devices are enough for determination of the future epicenter. The three devices will permit to research the correlation between Earth surface distribution of precursor function Sig and the Magnitude of the incoming earthquake.

The complex monitoring for solving the "when, where and how" earthquake prediction problem is very shortly discussed.